\documentclass[10pt,letterpaper]{article}
\usepackage[top=0.85in,left=2.75in,footskip=0.75in,marginparwidth=2in]{geometry}

\usepackage[utf8]{inputenc}

\usepackage{cite}

\usepackage{nameref,hyperref}

\usepackage{microtype}
\DisableLigatures[f]{encoding = *, family = * }

\raggedright
\usepackage{changepage}

\usepackage[aboveskip=1pt,labelfont=bf,labelsep=period,singlelinecheck=off]{caption}

\makeatletter
\renewcommand{\@biblabel}[1]{\quad#1.}
\makeatother

\usepackage{lastpage,fancyhdr,graphicx}
\usepackage{epstopdf}
\fancyheadoffset[L]{2.25in}
\fancyfootoffset[L]{2.25in}

\usepackage{color}

\definecolor{Gray}{gray}{.25}

\usepackage{graphicx}

\usepackage{sidecap}

\usepackage{wrapfig}
\usepackage[pscoord]{eso-pic}
\usepackage[fulladjust]{marginnote}
\reversemarginpar

\begin{document}
\vspace*{0.35in}

% title goes here:
\begin{flushleft}
{\Large
\textbf\newline{ Study of Stability and Consistency of EAS Thermal Neutron Detection at ENDA-64.}
}
\newline
% authors go here:
\\

% 作者信息
Heng-Yu Zhang\textsuperscript{1,2},
Xin-Hua Ma\textsuperscript{3,4,*},
Tian-Lu Chen\textsuperscript{1,2},
Shu-Wang Cui\textsuperscript{5},
Danzengluobu\textsuperscript{1,2},
Wei Gao\textsuperscript{3,4,*}
Wen-Chao Gao\textsuperscript{3,6},
Xin-Rui Gao\textsuperscript{1,2},
Zi-Ao Gong\textsuperscript{5},
Hai-Bing Hu\textsuperscript{1,2},
Denis Kuleshov\textsuperscript{7},
Kirill Kurinov\textsuperscript{7},
Bing-Bing Li\textsuperscript{8},
Fan-Ping Li\textsuperscript{9},
Jia-Heng Li\textsuperscript{10},
Yang Li\textsuperscript{5},
Hu Liu\textsuperscript{9},
Mao-Yuan Liu\textsuperscript{1,2,*},
Ye Liu\textsuperscript{11},
Xi-An Pan\textsuperscript{10},
Da-Yu Peng\textsuperscript{1,2},
Yao-Hui Qi\textsuperscript{5},
Dong Qu\textsuperscript{1,2},
Oleg Shchegolev\textsuperscript{7,12},
Yuri Stenkin\textsuperscript{7,12},
Tian-Shuang Yang\textsuperscript{11},
Li-Qiao Yin\textsuperscript{3,4},
Hui-Qian Zhang\textsuperscript{5},
Liang-Wei Zhang\textsuperscript{5},
Shi-Yuan Zhang\textsuperscript{5}

\bigskip
\bf{1} Key Laboratory of Cosmic Rays, Ministry of Education, Xizang University, Lhasa, Xizang 850000, China \\
\bf{2} Science School, Xizang University, Lhasa, Xizang 850000, China \\
\bf{3} Key Laboratory of Particle Astrophysics, Institute of High Energy Physics, Chinese Academy of Sciences, Beijing 100049, China \\
\bf{4} TIANFU Cosmic Ray Research Center, Chengdu, Sichuan 610000, China \\
\bf{5} College of Physics, Hebei Normal University, Shijiazhuang 050024, China \\
\bf{6} University of Chinese Academy of Sciences, Beijing 100049, China \\
\bf{7} Institute for Nuclear Research of the Russian Academy of Sciences, Moscow 117312, Russia \\
\bf{8} College of Physical Science and Technology, Bohai University, Jinzhou 121013, China \\
\bf{9} School of Physical Science and Technology, Southwest Jiaotong University, Chengdu, Sichuan 610031, China \\
\bf{10} School of Physics, Shandong University, Jinan 250100, China \\
\bf{11} School of Management Science and Engineering, Hebei University of Economics and Business, Shijiazhuang, Hebei 050061, China \\
\bf{12} Moscow Institute of Physics and Technology, 141700 Moscow, Russia
\\
\bigskip
* maxh@ihep.ac.cn,gaowei@ihep.ac.cn,liumy@utibet.edu.cn

\end{flushleft}

\section*{Abstract}
\textbf{Introduction:}Electron-Neutron Detector Array (ENDA) is designed to measure thermal neutrons produced by hadronic interactions between cosmic ray extensive air showers (EAS) and the surrounding environment as well as electrons around the cores of EAS. ENDA is located within Large High Altitude Air Shower Observatory (LHAASO). ENDA was expanded from an initial 16 detectors to 64 detectors in April 2023, so called ENDA-64, and has been running alongside LHAASO. The stability and consistency of neutron detection are crucial for laying a solid foundation for subsequent data analysis and physical results.

\textbf{Methods:} We obtain the stability by studying variations of event rate and thermal neutron rate in each cluster and the consistency by comparing distribution of number of thermal neutrons between clusters. Additionally, we investigate the specific influences of the rainy and dry seasons, as well as the presence or absence of sand cubes under the detectors, to examine the environmental factors affecting neutron measurement performance.

\textbf{Results:} The calibration results indicate good consistency in thermal neutron detection across the clusters, with the maximum inconsistency of 6.85\%. The maximum instability of event rate and thermal neutron rate over time are 4.68\% and 11.0\% respectively. The maximum inconsistency between the clusters without the sand cubes is 18\%. The use of sand cubes is effective in protecting the target material from rainwater, and the sand cubes help the cluster to increase collection of neutrons generated by EAS events.

% now start line numbers
%\linenumbers

% the * after section prevents numbering
\section{Introduction}
\hspace{2em}Cosmic ray energy spectrum is an important tool for investigating fundamental issues related to cosmic rays, such as origin, acceleration mechanism, and propagation in interstellar medium. 
The so-called "knee" region around $\sim$4 PeV, at which the spectral index changes from -2.7 to -3.1, was observed by several experiments 
 (e.g., \cite{ANTONI20051,AMENOMORI200658,Bartoli2015PRD,PhysRevLett.132.131002}). 
At high energy (above $10^{14}$ eV) cosmic ray energy spectrum measurement usually depended on the ground based experiments and mainly focused on energy determination and composition separation. The measurements of the "knee" still have deviations, in terms of the energy value of the knee, the magnitude and the power law indices of the flux before and after the knee. Individual composition measurement around "knee" can serve more powerful proofs to judgment of the various theoretical models proposed to explain the origin of the knee in cosmic ray spectrum (e.g., \cite{HORANDEL2004241,Gaisser2013,Nikolsky1997ICRC,Stenkin2011}). 

Most of ground based experiments measured electromagnetic component and muons in secondary particles of Extensive Air Showers (EAS) to obtain messages of primary cosmic rays, so as Large High Altitude Air Shower Observatory (LHAASO) \cite{Ma_lhaaso_cpc2022}, while the hadron components in EAS, which are typically concentrated at the core of EAS and carry important information about primary cosmic rays, have not been used to study cosmic ray energy spectrum and composition. Electron-Neutron Detector Array (ENDA) \cite{LiBB2021, PengDY2023, LiBB2024} was designed to measure secondary hadrons of EAS in conjunction with LHAASO, thereby enhancing the ability to determine energy and discriminate composition of the primary cosmic rays. 

When the secondary particles of EAS strike the array, the hadrons undergo nuclear reactions with the materials in the surrounding environment (such as soil, buildings, detector materials, and air), generating evaporation neutrons that are more than two orders of magnitude greater than the hadrons \cite{Stenkin2002MPLA,STENKIN2009293,Stenkin_2013}. The evaporation neutrons are then moderated to thermal neutrons by the materials in the surrounding environment.
The electron-neutron detector (END) can simultaneously measure the produced thermal neutrons and electrons in the EAS \cite{STENKIN2009293,Stenkin2007,STENKIN2008326}.  The first prototypes of such arrays were Multicom \cite{Djappuev2001ICRC} and PRISMA-32 \cite{Gromushkin2013}in Russia.

Later, the ENDs were firstly implemented at Yangbajing Cosmic Ray Observatory to study their performance in harsh high-altitude outdoor environments. The small prototype array named PRISMA-YBJ containing 4 ENDs was installed before inside the hall hosting of ARGO-YBJ \cite{BARTOLI201649}. 
Between PRISMA-YBJ and ARGO-YBJ, the coincident EAS events generated by cosmic rays were obtained, and a positive correlation between the thermal neutrons and the electromagnetic component  generated in EAS was confirmed. Besides, it was indicated that the ENDs can also be used to monitor seismic activities \cite{Stenkin2017,STENKIN2019105981}. 
Following PRISMA-YBJ, the prototype array of ENDA which contains 16 ENDs was decided to deploy in the LHAASO site \cite{LiBB2021}. At the same time, the PRISMA-YBJ was also extend to 16 ENDs \cite{LiBB_2017,LiuMY2020}. 
The ENDs of these two prototype arrays were upgraded and optimized, replacing the isotope element used in the scintillator for thermal neutron capture from $^{6}Li$ to $^{10}B$. %\textcolor{red}{(add sentences to illustrate why use 10B?)} 
Based on these two prototype arrays, a negative correlation between the thermal neutron counting rate and soil moisture was obtained ~\cite{LiuMY2020,Yang_2023}. 
Furthermore, the PRISMA-YBJ installed sand cubes under the detectors to study the influence of the target material, which is a major environmental factor affecting the arrays \cite{XiaoDX2022}. 
In order to achieve joint observations with LHAASO and ultimately energy spectrum in the knee region of cosmic rays using the main EAS component - hadronic one by recording thermal neutrons, ENDA has been expanded to 64 detectors so called ENDA-64 \cite{PengDY2023,LiBB2024}. The influence of "sand cubes" was reexamined in this work. And the measurements of neutrons were monitored to assess the performance of the array.

\section{Experimental Setup}\label{sec2}
\begin{figure}[ht]
\centering

% 左侧子图 (a)
\begin{minipage}{0.49\textwidth}
\centering
\includegraphics[width=\linewidth]{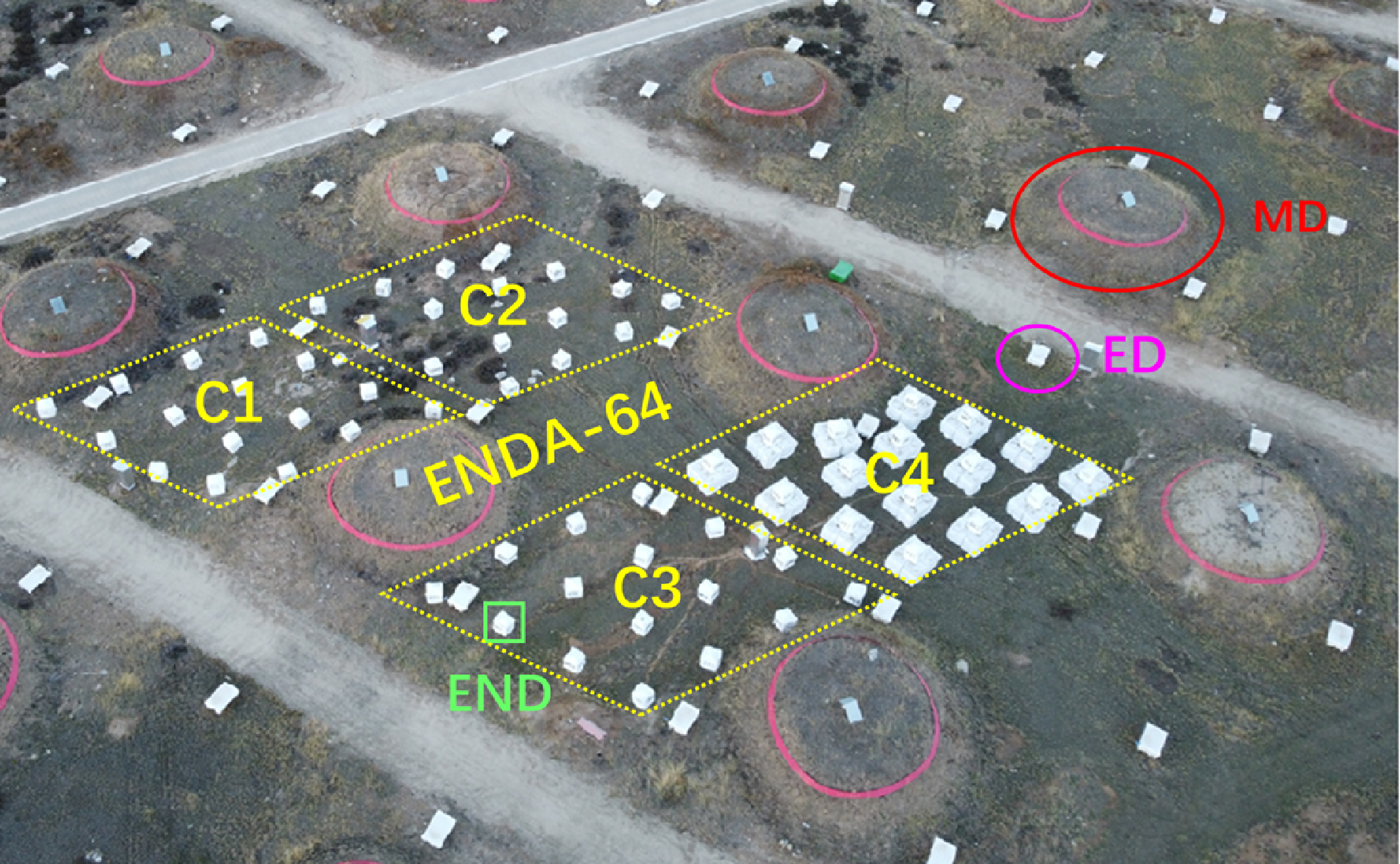}
\end{minipage}
\hfill
% 右侧子图 (b)
\begin{minipage}{0.50\textwidth}
\centering
\includegraphics[width=\linewidth]{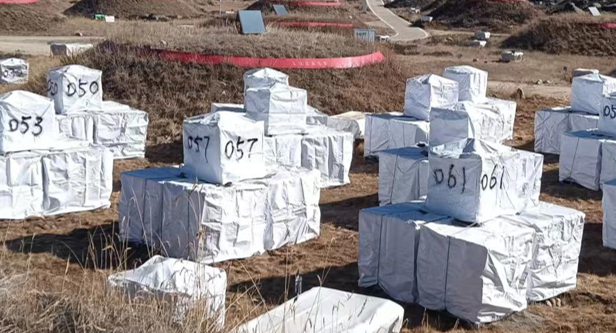}
\end{minipage}

\caption{\color{Gray} \textbf{} (a) Aerial view of ENDA-64 within the LHAASO Array. ENDs and four clusters of ENDA and electron detectors (EDs) and muon detectors (MDs) of LHAASO are marked. (b) ENDs of cluster \#4 with sand cubes.}
\label{fig:sandcube_photo}

\end{figure}

\hspace{2em}ENDA-64 locates inside one km$^2$ Array (KM2A), the sub-array of LHAASO. It contains 4 clusters and each cluster consists of 16 detectors, as shown in Fig. \ref{fig:sandcube_photo}. Each cluster's detectors are arranged in a 4 $\times$ 4 configuration, with a distance of 4.5 meters between adjacent detectors, shaping a diamond pattern.
After the ENDA-64 operating a period, cluster No. 4 was selected to build the sand cubes to study the influence of surround materials. As shown in Fig. \ref{fig:sandcube_photo}, four cuboid plastic boxes with size of 1 m $\times$ 1.2 m $\times$ 1.15 m, filled with sand with height of 1 m, are installed beneath each detector.

The detector mainly composed of scintillator, scintillation light collecting cone, Photomultiplier Tube (PMT), the front-end electronics (FEE), and black housing. The scintillator is made of ZnS(Ag) and B\textsubscript{2}O\textsubscript{3} alloy, deposited in optical silicone rubber. Natural Boron contains $19\%$ of $^{10}$B. The effective thickness of the scintillator layer is 50 $mg/cm^{2}$. 
The following equation illustrates the principle of neutron capture, where the isotope $^{10}B$ undergoes a nuclear reaction with thermal neutrons.
\begin{equation}
\label{eq:boron_li}
\begin{array}{l}
^{10}\mathrm{B} + n \rightarrow ^{7}\mathrm{Li}^* + \alpha \\
\quad \rightarrow ^{7}\mathrm{Li} + \alpha + \gamma\,(0.48\,\mathrm{MeV}) + 2.3\,\mathrm{MeV}\,(94\%) \\
\quad \rightarrow ^{7}\mathrm{Li} + \alpha + 2.8\,\mathrm{MeV}\,(6\%)
\end{array}
\end{equation}

\begin{figure}[ht]
\centering
\includegraphics[width=0.4\linewidth]{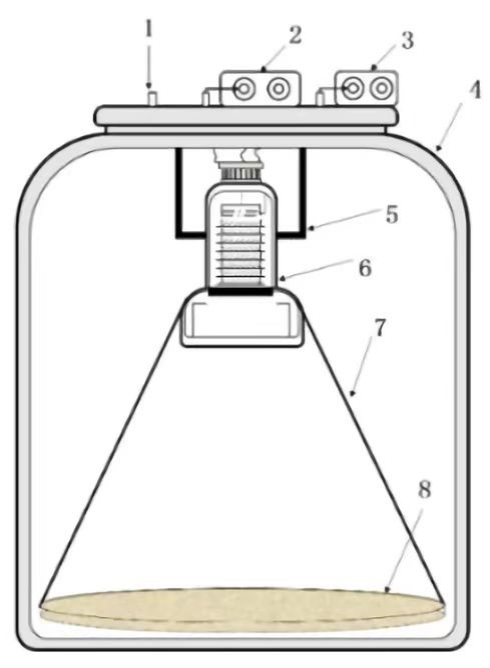}
\caption{\textbf{Schematic diagram of the END.} 1 - the high-voltage input port, 2 - DIU connected to the 8th dynode of the PMT, 3 - IU connected to the 5th dynode of the PMT, 4 - a black tank for the detector housing, 5 - the PMT fixed holder, 6 - the PMT, 7 - the scintillation light collecting cone, and 8 - the scintillator. \cite{DONG2019162639}.}
\label{fig:detector_photo}
\end{figure}

Fig. \ref{fig:detector_photo} shows the schematic diagram of the END.  
The scintillator is positioned at the bottom of a cylindrical polyethylene (PE) tank, which serves as the detector housing. A 4-inch PMT (Beijing Hamamatsu, model CR-165) is mounted on the lid of the tank, positioned 0.3 m away from the scintillator.
A conical reflective layer is placed between the scintillator and the PMT to enhance the collection of scintillation photons.

The FEE of each detector consists of a voltage divider for PMT, a discriminator-integrator unit (DIU) and an integrator unit (IU). The DIU connects to the 8th dynode of the PMT for measuring energy deposit measurements, selecting coincidences, and counting the number of secondary neutrons. Meanwhile, the IU connects to the 5th dynode of the PMT to extend its dynamic range. 
The output signals from the END include both weak but fast signals generated by charged particles and high-amplitude, slow, and delayed signals produced from captured thermal neutrons.

A data acquisition system (DAQ) to each cluster consists of a 32-channel flash analog-to-digital converter (FADC) connected to a PC via optical fiber. The first 16 channels of the FADC are used for receiving the signals from the 8th dynode of the PMTs and the rest 16 for the 5th dynode. The first pulse produced mostly by the EAS charged particles, is used as a trigger signal and for energy deposit measurements, while the delayed neutron capture pulses are counted within a time gate of 20 ms to determine the number of detected thermal neutrons. If any 2 of the 16 detectors coincide within a window of 1 µs (trigger type M1), the FADC performs sampling. This process constitutes the first-level triggering. The online program then analyzes the received input signals and perform the second-level triggering as follows ~\cite{LiBB_2017,PengDY2023}.
\begin{enumerate}
  \item At least 2 detectors must produce a first-level trigger, with signals of at least 10 minimum ionization particles (m.i.p.s).
  \item The total energy deposit power measured is larger than 125 m.i.p.s. 
  \item The total number of thermal neutrons recorded is at least 3.
\end{enumerate}
If condition 1 is satisfied, the trigger type is M1; if both conditions 1 and 2 are satisfied, the trigger type is M2; if both conditions 1 and 3 are satisfied, the trigger type is M3; and if conditions 1, 2, and 3 are all satisfied, the trigger type is M4. Additionally, the online program generates a software trigger signal every minute to count number of neutrons within a 20 ms time window for measuring background of neutrons (M0 events).

\section{Results}

\begin{figure}[ht]  % Using arXiv's recommended placement specifier
\centering
\includegraphics[width=0.5\linewidth]{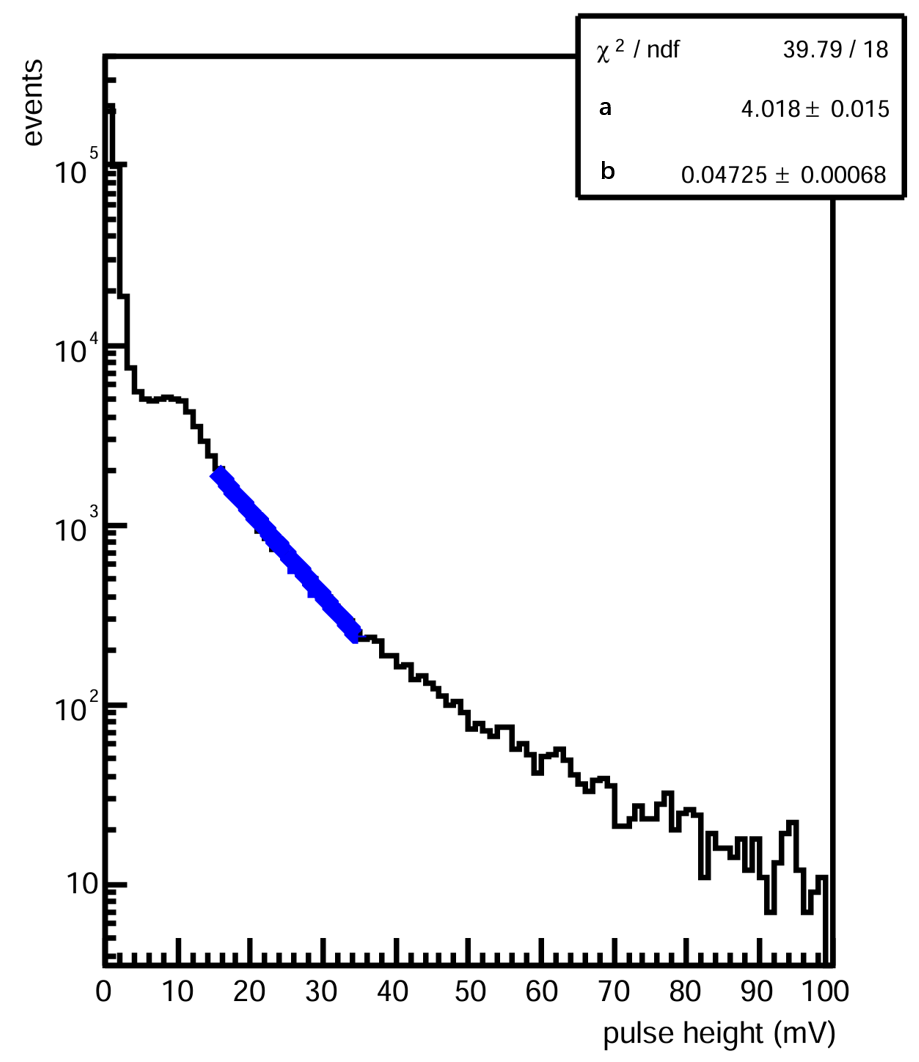}  % Changed \textwidth to \linewidth
\caption{\textbf{Pulse height distribution.} Black line shows the measurement from one detector; blue line represents the fitting function (Eq.~\ref{eq:calibration}).}  % Added \textbf, improved formatting
\label{fig:Calibration}
\end{figure}

\begin{table}[htp]
\centering
\renewcommand{\arraystretch}{2.5} % 调整行高
\caption{$M$, $S$ and $R$ of steepness ($b$) of the pulse height distribution.}
\begin{tabular}{|c|c|c|c|c|}
\hline
%\rowcolor[HTML]{4472C4} % 蓝色背景
%\textcolor{white}{\textbf{}} & 
cluster No. & 1 & 2 & 3 & 4 \\ \hline
$M (10^{-2}$) & 4.91 & 4.66 & 4.78 & 5.01 \\ \hline
$S (10^{-3}$) & 1.36 & 3.19 & 2.99 & 1.50 \\ \hline
$R (\%)$      & 2.77 & 6.85  & 6.26  & 2.99  \\ \hline
\end{tabular}
\label{tab:Calibration}
\end{table}

\subsection{Calibration of neutron detection}

\hspace{2em}First of all, each detector should be calibrated for neutron detection. When one $^{10}$B captures a thermal neutron, charged nuclei ($\alpha$ and $^7Li$) are generated, which then pass through the scintillator, depositing certain amounts of energy to produce photons, as shown in Eq. \ref{eq:boron_li}. 
Part of the photons are collected by the light cathode of PMT to generate an electrical pulse. The pulse height corresponding to a single thermal neutron should be the same across all detectors by adjusting the high voltage values on the PMTs. In practice, this is achieved using the pulse height distribution.
As an example, Fig. \ref{fig:Calibration} shows the pulse height distribution of one detector. The peak at x=0 is a noise spectrum produced in events when common start FADC trigger was produced by one of another 15 detectors. Next wide maximum at about 10 mV is produced by trigger threshold of this detector. The distribution consists of three parts: the low part primarily includes noise, the middle part corresponds to a single thermal neutron, and the high part is produced by fast CR neutrons and proton background through recoil protons in silicone. 

The fitting function in the middle part follows a power law:
\begin{equation}
\label{eq:calibration}
y = 10^{a-bx}
\end{equation}
where $a$ quantifies a statistical parameter of the distribution, and $b$ indicates the steepness of the distribution. $b$ reflects the energy deposited by a single thermal neutron in the scintillator; thus, it is expected that $b$ should ideally have the same value across all detectors used in this experiment.

After adjusting the high voltage values on the PMTs, the pulse height distributions of all detectors in each cluster are measured %(see Fig. \ref{fig:Calibration-1}). 
The mean values ($M$) and standard deviations ($S$) of $b$ for each cluster are then obtained and presented in Table \ref{tab:Calibration}. It is shown that after calibration, the ratio ($R$) of $S$ to $M$, which represents the inconsistency in single thermal neutron detection among detectors in each cluster, reaches the maximum of $6.85\%$.

\begin{figure}[ht]
\centering

% 上方的图
\begin{minipage}[b]{\textwidth}
\centering
\includegraphics[width=0.6\linewidth]{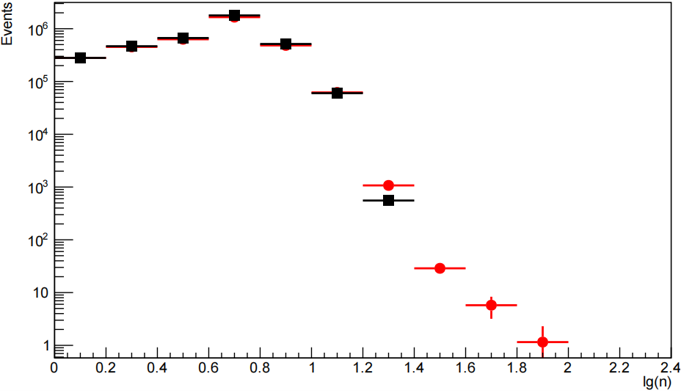}
\vspace{0.5em} % 添加小间距
\end{minipage}

% 下方的图
\begin{minipage}[b]{\textwidth}
\centering
\includegraphics[width=0.6\linewidth]{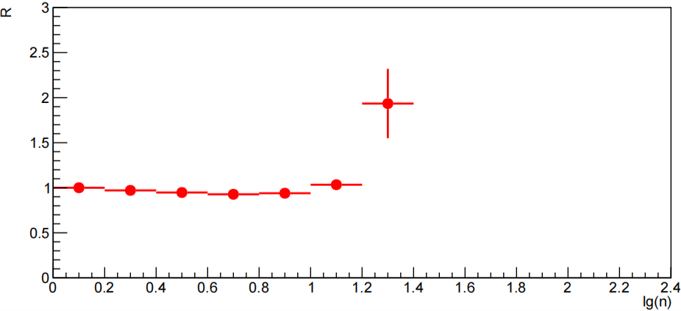}
\end{minipage}

\caption{\textbf{Thermal neutron distributions.} \textbf{Upper panel:} Number of thermal neutrons in one cluster from February 9 to June 3, 2024. Red dots show triggered events; black squares show M0 background events. \textbf{Lower panel:} Ratio ($R$) of triggered events to M0 background events.}
\label{fig:trig-bg-C02}
\end{figure}

\subsection{Distribution of number of thermal neutrons} 

\hspace{2em}The differential distribution of the number of thermal neutrons is the key measurement of thermal neutrons in EAS events. For example, data from one cluster during the period from Fabruary 9, 2024, to June 3, 2024, are used to obtain the distributions of both triggered events and M0 background events (Fig. \ref{fig:trig-bg-C02}).
The distribution of M0 background events is normalized to that of triggered events by using event rates. 
The ratio of the distribution of triggered events to that of M0 background events shows that at lower neutron counts ($lg(n) < 1$), $R$ is less than 1.1, indicating that the triggered events mainly consist of background events. In contrast, at higher neutron counts ($lg(n) > 1$), $R$ is greater than 1.2, indicating that the triggered events are primarily real EAS events. The same results are obtained for the other three clusters.

\begin{figure}[ht]  % Using arXiv's recommended [ht] placement
\centering
\includegraphics[width=0.6\linewidth]{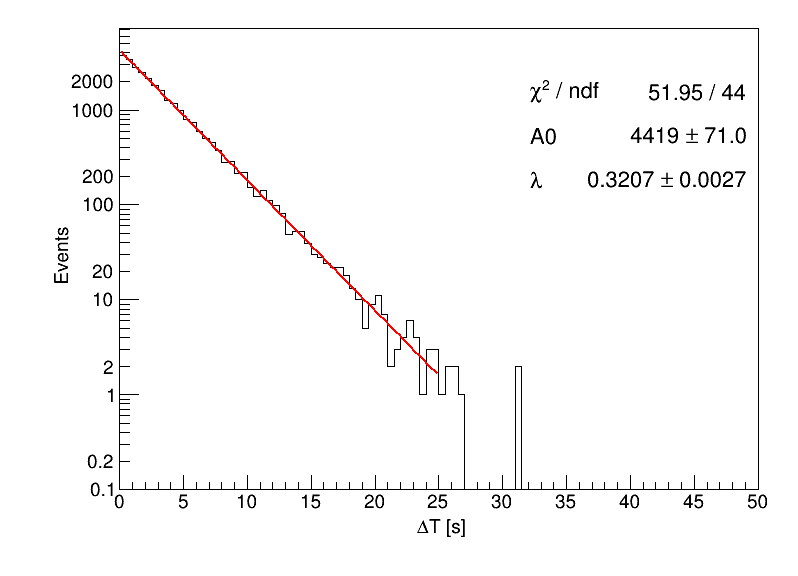}  % Changed \textwidth to \linewidth
\caption{\textbf{Time difference distribution.} Black line shows time differences between adjacent events in one cluster during one day; red line represents the fitting function (Eq.~\ref{eq:lifetime}).}
\label{fig:timediffdist}
\end{figure}

\begin{figure}[ht]  % Using arXiv's recommended placement
\centering
\includegraphics[width=0.6\linewidth]{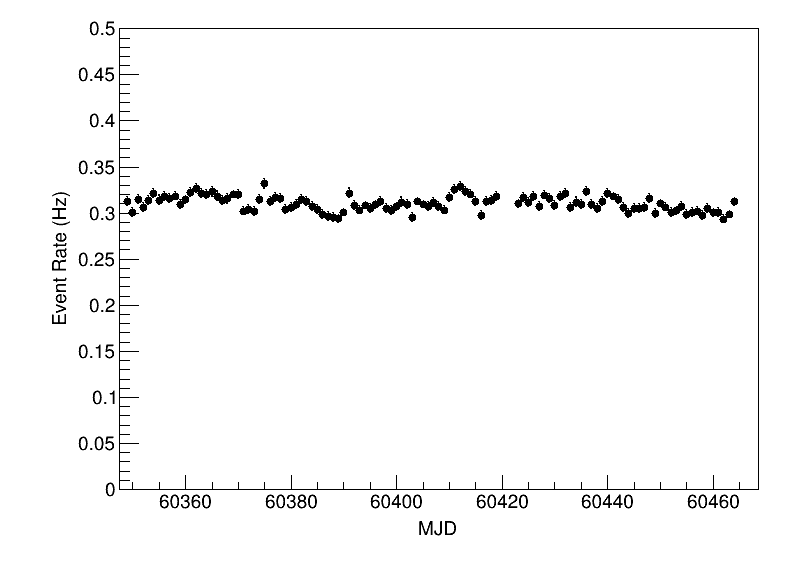}  % \linewidth instead of \textwidth
\caption{\textbf{Event rate variation.} Measured in one cluster from February 9 to June 3, 2024 (MJD: Modified Julian Day).}
\label{fig:eventrate}
\end{figure}

\begin{table}[htp]
\centering
\renewcommand{\arraystretch}{2.5} % 调整行高
\caption{$M$, $S$ and $R$ of event rates.}
\begin{tabular}{|c|c|c|c|c|}
\hline
cluster No. & 1 & 2 & 3 & 4 \\ \hline
$M (10^{-1}Hz) $ & 3.16 & 3.42 & 3.10 & 3.15  \\ \hline
$S (10^{-3}Hz) $ & 1.30 & 1.60 & 8.22 & 1.23  \\ \hline
$R (\%)        $ & 4.11 & 4.68 & 2.65 & 3.90  \\ \hline
\end{tabular}
\label{tab:EventRates}
\end{table}

% 插入表格
\begin{table}[htp]
\centering
\renewcommand{\arraystretch}{2.5} % 调整行高
\caption{$M$, $S$ and $R$ of thermal neutron rates.}
\begin{tabular}{|c|c|c|c|c|}
\hline
cluster No. & 1 & 2 & 3 & 4 \\ \hline
$M (Hz) $ & 1.34  & 1.48  & 1.19  & 1.25  \\ \hline
$S (Hz) $ & 0.147 & 0.127 & 0.067 & 0.118 \\ \hline
$R (\%) $ & 11.0  & 8.58  & 5.63  & 9.44  \\ \hline
\end{tabular}
\label{tab:NeutronRates}
\end{table}

\subsection{Instability and inconsistency} 

\hspace{2em}It is important to estimate the systematic uncertainty of the array by measuring instability and inconsistency using real data.  Event rate in one cluster in one day is obtained from time difference distribution between adjacent events (Fig. \ref{fig:timediffdist}), fitted by an exponential function:
\begin{equation}
\label{eq:lifetime}
y = A_0e^{-{\lambda}x}
\end{equation}
where $A_0$ quantifies statistical parameter of the distribution, and $\lambda$ (Hz) is event rate. For each cluster in period from February 9, 2024 to June 3, 2024, variation of event rate (Fig. \ref{fig:eventrate}) and then M, S and R of event rate of each cluster are obtained (Tab. \ref{tab:EventRates}). It can be seen that the maximum instability of the event rate in four clusters is 4.68\%. Furthermore, in the same way, for each cluster, variation of thermal neutron rate and then M, S and R of the thermal neutron rate are obtained (Tab. \ref{tab:NeutronRates}). The maximum instability of thermal neutron rate in four clusters is 11.0\%.

\begin{figure}[ht]
\centering

% Upper panel
\begin{minipage}[b]{\textwidth}
\centering
\includegraphics[width=0.6\linewidth]{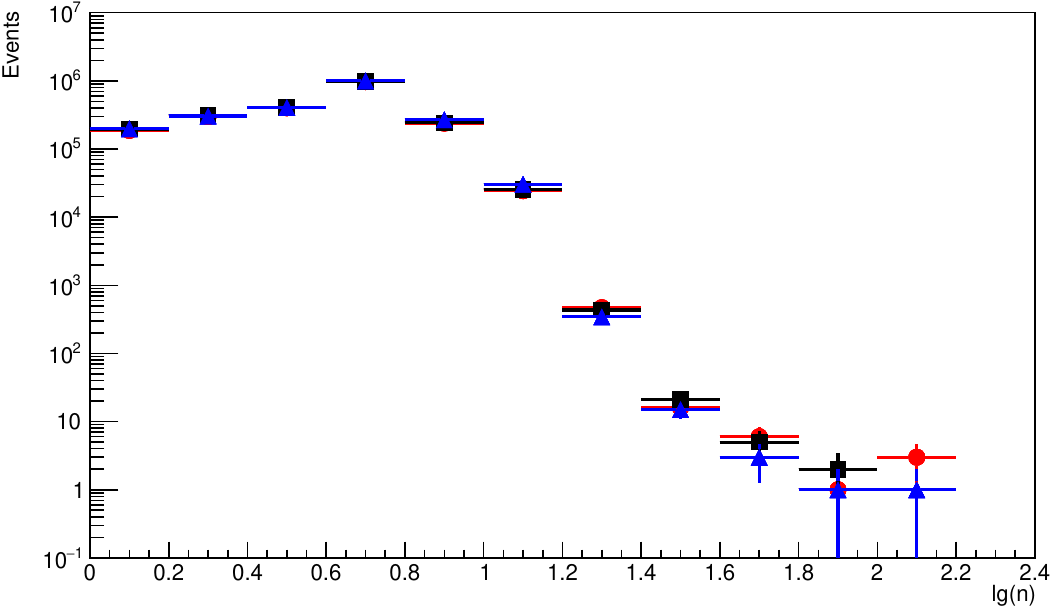}
\vspace{0.5em} % Small spacing between figures
\end{minipage}

% Lower panel
\begin{minipage}[b]{\textwidth}
\centering
\includegraphics[width=0.6\linewidth]{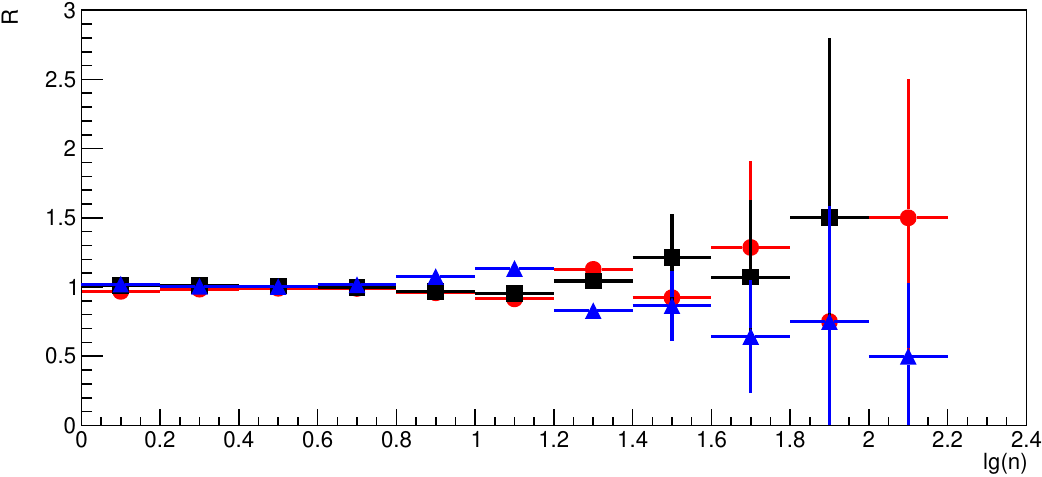}
\end{minipage}

\caption{\textbf{Thermal neutron consistency analysis.} \textbf{Upper panel:} Distributions of thermal neutrons from triggered events in Cluster 1 (black squares), Cluster 2 (red dots), and Cluster 3 (blue triangles) during August 15 to December 1, 2024. \textbf{Lower panel:} Ratios (R) of each cluster's distribution to their average values.}
\label{fig:Consistency without C04}
\end{figure}

In order to study inconsistency, data from the same period—August 15, 2024, to December 1, 2024—from clusters No. 1, 2, and 3 without sand cubes are used to compare the distributions (Fig. \ref{fig:Consistency without C04}). The results show that the maximum difference among the three clusters is 18\%, disregarding the higher neutron counts, which have low statistics and consequently large statistical errors.

\subsection{Effect of sand cubes}
 
\hspace{2em}Sand cubes were expected to protect the target material from rainwater, which can reduce the number of thermal neutrons reaching the detectors. Moreover, the volume of the sand cubes is 4 $m^3$ to increase the geometrical factor, which is occupancy ratio of target material (soil, sand, etc.) in air,  close to 1, ensuring that the number of collected neutrons is comparable to that detected by the instruments mounted on the ground. 
In order to assess the effect of the sand cubes, two comparisons are performed. 

\begin{figure}[ht]
\centering

% Upper panel
\begin{minipage}[b]{\textwidth}
\centering
\includegraphics[width=0.6\linewidth]{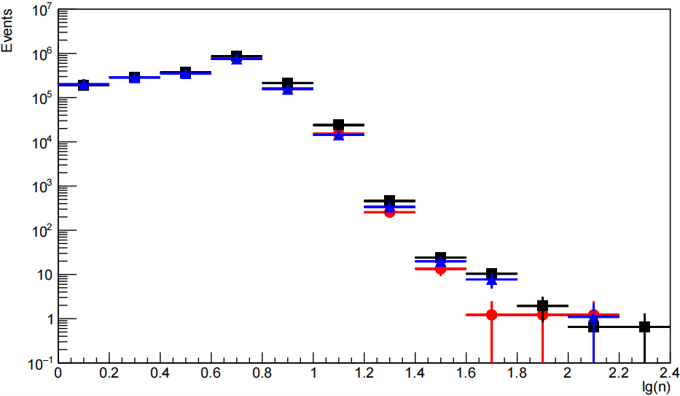}
\vspace{0.5em} % Small spacing between figures
\end{minipage}

% Lower panel
\begin{minipage}[b]{\textwidth}
\centering
\includegraphics[width=0.6\linewidth]{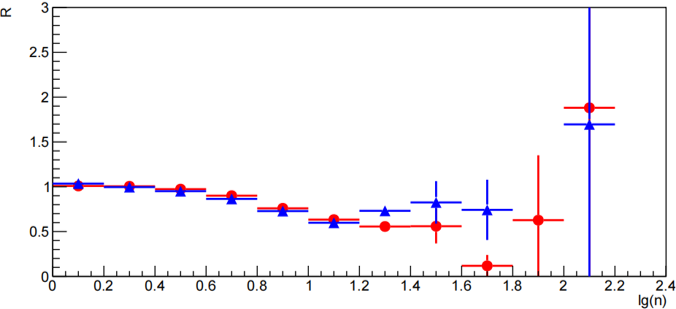}
\end{minipage}

\caption{\textbf{Thermal neutron distributions with/without sand cubes.} 
\textbf{Upper Panel:} Thermal neutron counts from triggered events in Cluster 4 during: (1) dry season with sand cubes (Feb 9-Jun 3, 2024; black squares), (2) rainy season without sand cubes (Aug 16-Oct 14, 2023; red dots), and (3) rainy season with sand cubes (Jun 4-Aug 14, 2024; blue triangles). 
\textbf{Lower Panel:} Ratios ($R$) of rainy season distributions (with/without cubes) relative to dry season baseline.}
\label{fig:Sandcube_Ratio2}
\end{figure}

Firstly, a comparison is made among the distributions in cluster No. 4 over three periods: (1) during the dry season with sand cubes; (2) during the rainy season without sand cubes; and (3) during the rainy season with sand cubes (Fig. \ref{fig:Sandcube_Ratio2}).
Ratios ($R$) of the two distributions during the rainy season to those during the dry season are calculated individually.
At lower neutrons ($lg(n)<1.2$), the two ratios are similar within the statistical error, which is attributed to instability during the periods. At higher neutrons ($lg(n) > 1.2$), the fact that the ratio of the distribution during the rainy season with sand cubes is higher than that of the distribution during the rainy season without sand cubes indicates that the sand cubes do indeed play a role in protecting the target material from rainwater. 

\begin{figure}[ht]
\centering

% Upper panel
\begin{minipage}[t]{\textwidth}
\centering
\includegraphics[width=0.6\linewidth]{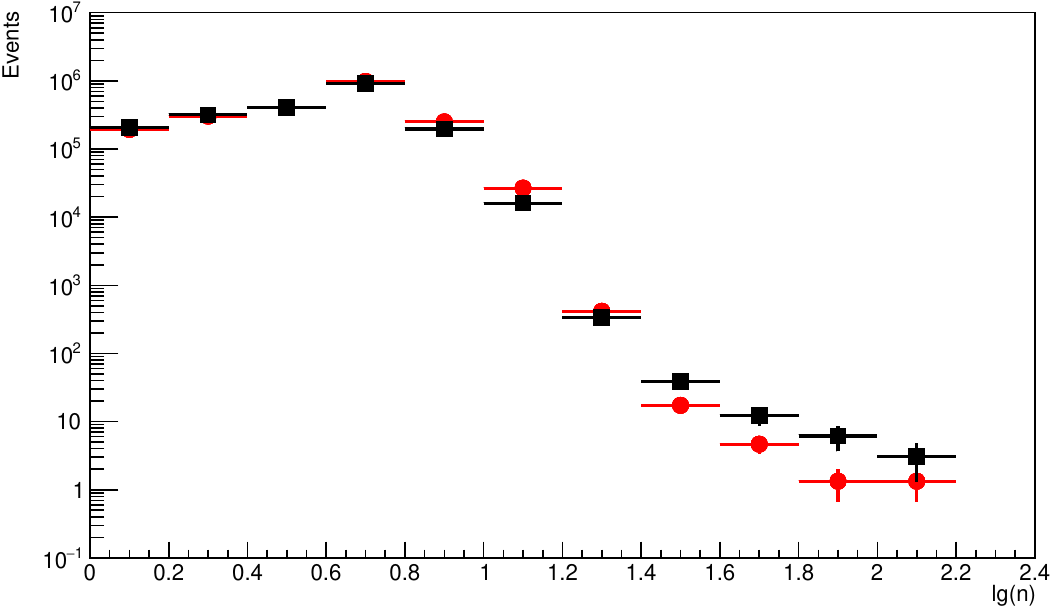}
\par\vspace{1ex} % Vertical spacing between figures
\end{minipage}

% Lower panel
\begin{minipage}[t]{\textwidth}
\centering
\includegraphics[width=0.6\linewidth]{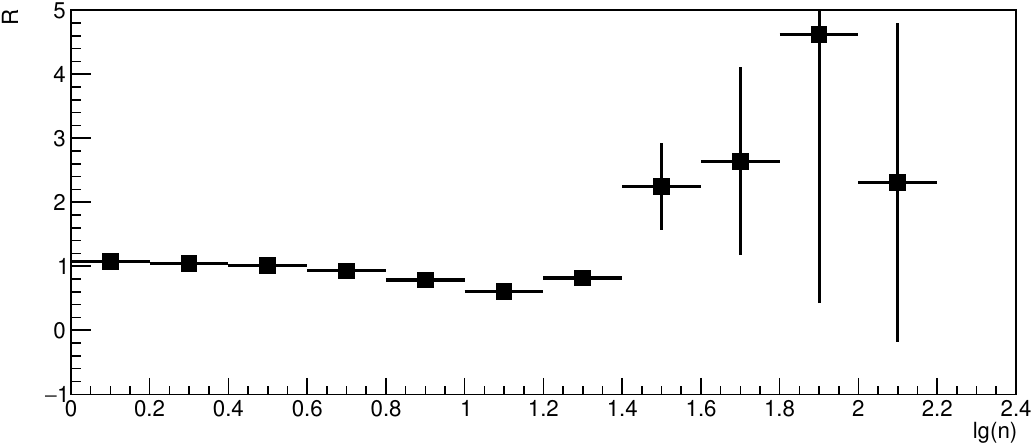}
\end{minipage}

\caption{\textbf{Thermal neutron comparison with sand cubes.} 
\textbf{Upper Panel:} Distribution of thermal neutron counts from triggered events in Cluster 4 (black squares) versus average distribution in other clusters (red dots). 
\textbf{Lower Panel:} Ratio ($R$) of Cluster 4 distribution to average of other clusters.}
\label{fig:Sandcube_Ratio}
\end{figure}

Secondly, a comparison is made between the distribution in cluster No. 4 with sand cubes and the average distribution in the other three clusters without sand cubes during the dry season (Fig. \ref{fig:Sandcube_Ratio}) as shown in the same period in Fig. \ref{fig:Consistency without C04}.
Along $lg(n)$ from 0, the ratio of the two distributions decreases from 1 to 0.5. After $lg(n) > 1.2$, it rises above 1, although there statistics is poor. This indicates that the cluster with sand cubes has a lower background neutron count than the one without sand cubes at $lg(n) < 1.2$, which can be explained by the lower radioactivity in the sand cubes compared to the soil. 
At $lg(n) > 1.2$, the cluster with sand cubes captures more thermal neutrons due to EAS events than the one without sand cubes, because the cluster with sand cubes is 1 meter above the ground, allowing more thermal neutrons to come from a wider area through the atmosphere than in the cluster without sand cubes.

\section{Conclusion}
In this study, we investigate the performance of ENDA-64 in detecting thermal neutrons generated by EAS. The calibration, instability, and inconsistency of neutron detection, as well as the effects of environmental factors such as seasonal variations and the presence of sand cubes, have been thoroughly examined. 
The calibration results indicate good consistency in thermal neutron detection across the clusters, with the maximum inconsistency of 6.85\%, which can be reduced by more careful adjustment of high voltage values of all the detectors. The maximum instability of event rate and thermal neutron rate are 4.68\% and 11.0\% respectively which were mainly from noise in the clusters and noise has been reduced since the periods. The maximum inconsistency between the clusters without the sand cubes is 18\% which can be also reduced when the calibration is done more carefully. Effect of instability and inconsistency on the systematic uncertainties of the final cosmic ray energy spectrum and composition is expected through energy estimation and component separation by using neutrons. It can be obtained by using the simulation method which will be done in the near future.
The use of sand cubes in cluster No. 4 is effective in protecting the target material from rainwater, and the sand cubes help the cluster to increase collection of neutrons generated by EAS events. Therefore, the sand cubes can increase neutrons collection especially at lower energy so as to decrease energy threshold to give more accurate measurement of cosmic ray energy spectrum and better component separation at lower energy range. It will be studied both by using more real data and with simulation method.
This study gives fundamental dependence for improvement of performance of the array, and provides a strong foundation for analysis of cosmic ray energy spectrum and composition measurement by using ENDA. After getting enough statistics, the ENDA-64 will give results of measurement of cosmic ray energy spectrum around the knee region of light component (proton and Helium). After it, the array is planned to be extended to the ENDA-400 composed by 400 EN-detectors to measure the  knee region of heavy components (Fe). Also, we plan to modify the data acquisition making it able to study continuously background thermal neutron variations. 

\clearpage
\section*{Acknowledgments}

This work was supported in China by the National Natural Science Foundation of China (NSFC, No.12320101005, No.12263005, No.12373105 and No.12205244) Natural Science Foundation of the Tibet Autonomous Region (grant No.XZ202501ZR0058). The authors would like to thank all staff members who work year-round at the LHAASO site, which is located at an altitude of 4,400 meters above sea level, for maintaining the detectors and ensuring the electricity supply and other components of the experiment run smoothly.

%\nolinenumbers

%This is where your bibliography is generated. Make sure that your .bib file is actually called library.bib
\bibliography{main.bib}

%This defines the bibliographies style. Search online for a list of available styles.
\bibliographystyle{unsrt}

\end{document}